\documentclass[%
 reprint,
 floatfix,
superscriptaddress,
nofootinbib,
 amsmath,amssymb,
 aps,
prd,
showkeys,
]{revtex4-2}

\usepackage{graphicx}
\usepackage{dcolumn}
\usepackage{bm}
\usepackage{hyperref}
\usepackage[mathlines]{lineno}
\usepackage{epsfig}
\usepackage{float}
\usepackage{amssymb}
\usepackage{multirow}
\usepackage{wrapfig}
\usepackage{array}
\usepackage[dvipsnames]{xcolor}
\usepackage{amsmath}
\usepackage{natbib}
\usepackage{ulem}
\usepackage{multirow}
\usepackage{cleveref}
\usepackage{slashed} 
\crefname{equation}{equation}{equations}
\usepackage{booktabs}
\usepackage{subfig}
\newcommand*\mnras{MNRAS}
\newcommand*\aap{A\&A}
\newcommand*\aj{AJ}

\newcommand*\jcap{J. Cosmology Astropart. Phys.}

\begin{document}
\normalem

\preprint{APS/123-QED} 

\title{Runaway dilaton models: Improved constraints from the full cosmological evolution}

\author{L\'eo Vacher}
\email{leo.vacher@irap.omp.eu}
\affiliation{Institut de Recherche en Astrophysique et Planétologie, CNRS, CNES, 31028 Toulouse, France\\}
\affiliation{Universit\'e de Toulouse UPS, 31028 Toulouse, France\\}
\author{Nils Sch\"oneberg}%
\email{nils.science@gmail.com}
\affiliation{Dept. F\'isica Qu\`antica i Astrof\'isica, Institut de Ci\`encies del Cosmos (ICCUB), Facultat de F\'isica, Universitat de Barcelona (IEEC-UB), Mart\'i i Franqu\'es, 1, E08028 Barcelona, Spain\\}%
\author{J. D. F. Dias}
\affiliation{Centro de Astrof\'{\i}sica da Universidade do Porto, Rua das Estrelas, 4150-762 Porto, Portugal}%
\affiliation{Instituto de Astrof\'isica e Ci\^encias do Espa\c co, CAUP, Universidade do Porto, Rua das Estrelas, 4150-762, Porto, Portugal}
\affiliation{Faculdade de Ci\^encias, Universidade do Porto, Rua Campo Alegre, 4169-007, Porto, Portugal}
\author{C. J. A. P. Martins}%
\email{Carlos.Martins@astro.up.pt}
\affiliation{Centro de Astrof\'{\i}sica da Universidade do Porto, Rua das Estrelas, 4150-762 Porto, Portugal}%
\affiliation{Instituto de Astrof\'isica e Ci\^encias do Espa\c co, CAUP, Universidade do Porto, Rua das Estrelas, 4150-762, Porto, Portugal}
\author{Francisco Pimenta}%
\affiliation{Centro de Astrof\'{\i}sica da Universidade do Porto, Rua das Estrelas, 4150-762 Porto, Portugal}%
\affiliation{Faculdade de Ci\^encias, Universidade do Porto, Rua Campo Alegre, 4169-007, Porto, Portugal}
\date{\today}

\begin{abstract}
One of the few firm predictions of string theory is the existence of a massless scalar field coupled to gravity, the dilaton. In its presence, the value of the fundamental constants of the universe, such as the fine-structure constant, will vary with the time-dependent vacuum expectation value of this field, in direct violation of the Einstein equivalence principle. The \emph{runaway dilaton} proposed by Damour, Piazza, and Veneziano provides a physically motivated cosmological scenario which reconciles the existence of a massless dilaton with observations, while still providing nonstandard and testable predictions. Furthermore, the field can provide a natural candidate for dynamical dark energy. While this model has been previously constrained from local laboratory experiments and low-redshift observations, we provide here the first full self-consistent constraints, also including high redshift data, in particular from the cosmic microwave background. We consider various possible scenarios in which the field could act as quintessence. Despite the wider parameter space, we make use of recent observational progress to significantly improve constraints on the model, showing that order unity couplings (which would be natural in string theory) are ruled out.
\end{abstract}

\keywords{Cosmology, varying constants, dark energy, string theory, scalar fields, modified gravity} 
\maketitle
\section{\label{sec:intro} Introduction}

The discovery of the Higgs boson at the LHC \citep{Higgs1,Higgs2}, confirmed that spin-0 scalar fields are part of the building blocks of nature. As they are easy to couple to gravity without breaking covariance, they are now commonly invoked as a powerful tool to model cosmological paradigms, including quintessence, early dark energy, inflation, symmetry breaking phase transitions (with their associated topological defects), and---last but not least---dynamical varying couplings \cite{Martins2017review}.

Moreover, they appear as a theoretical necessity in most of grand unification scenarios and attempts of building a quantum theory of gravity.
This is the case of string theory, one of the most promising paths connecting quantum field theories and gravity (for a review see e.g. \cite{Stringreview}). 
Indeed, many bridges have already been built between gravity and quantum fields thanks to quantum strings, such as the recent AdS$-$CFT correspondence and similar applications of the holographic principle (see e.g. \cite{Susskind}). Even though it is still impossible to tell what the final form of the theory should be, one of its uncircumventable predictions seems to be the existence of a  scalar partner to the graviton field, called the \emph{dilaton}. Its dynamics sets the intensity of the various interactions between strings through the \emph{string coupling}, and therefore that of the fundamental forces of the standard model. Among other things, the evolution of the dilaton field implies a variation of all the fundamental dimensionless couplings, such as the fine-structure constant. In turn, this implies a violation of the Einstein equivalence principle \cite{Will2014,Will2017}). 

Theory suggests that the dilaton should be massless, which would be in violent contradiction with observations. To overcome such an issue in a physically motivated manner, it has been proposed that the dilaton coupling to other matter fields is attracted toward finite smooth limits \citep{DamourPolyakov1994,DamourNordtvedt1993,DamourPiazzaVeneziano2002}. This model is called the \emph{runaway dilaton} and has the advantage of providing clear predictions, that can be confronted with observations. As such, it can be used as a very compelling testbed model 
to implement and study variations of fundamental constants on cosmological scales. Moreover, with a suitable choice of potential $V(\phi)$ or extra couplings, the dilaton field can provide a physically motivated source of dynamical dark energy \cite{Gasperini2001}.

The present work builds upon several previous phenomenological studies \citep{Gasperini2001,MartinsVielzeuf2015,Martinelli2015,MartinsVacher2019} while aiming to be more accurate and more general. This is achieved by confronting the full cosmological field evolution with the latest datasets, as done in \cite{Vacher2022} for Bekenstein models, while freeing ourselves from assumptions made in previous studies. In \cref{sec:pheno} we introduce the evolution equations of the coupled dilaton field, as well as their impact on various observables. In \cref{sec:datasets} we present the datasets we use in order to obtain the constraints discussed in \cref{sec:results}. Finally, we present our conclusions in \cref{sec:conclusion}.

\section{\label{sec:pheno} Phenomenology of the coupled runaway dilaton}

The dilaton field $\Phi$ appears in every string and superstring theory as a massless scalar excitation of the bosonic string. It comes as a massless scalar mode on the first exited state of the closed string along with two rank-2 tensor fields: the symmetric metric tensor $\tilde{g}_{\mu \nu}$ and the antisymmetric Neveu-Schwarz $B$-field $B_{\mu \nu}$, which plays a role comparable to an electromagnetic gauge field for extended objects. As such, $\Phi$ is a partner of the graviton and contributes to the behavior of gravity itself (for an elementary introduction see e.g. \cite{SzaboString}).
At tree level, it is expected to be coupled to the various sectors in the string-frame Lagrangian through coupling functions $B_i(\Phi)$ with $i = \tilde{g},F,\psi,\Phi ...$
While string theory cannot predict the exact form of these coupling functions, the assumption underlying the \emph{runaway dilaton} model is that they can naturally be attracted toward a finite smooth limit \cite{DamourNordtvedt1993} as
\begin{equation}
    B_i(\Phi)= C_i + \mathcal{O}(e^{-\Phi})\,.
\label{eq:barecoupling}
\end{equation}
This can reconcile a massless dilaton with experimental observations while still providing many nonstandard but observable predictions.

The direct coupling of $\Phi$ to gravity is reabsorbed in a conformal transformation of the metric $\tilde{g}\to g$ and a redefinition of the field $\Phi \to \phi$ \citep{DamourPolyakov1994}, leading to an effective low energy Lagrangian density in the Einstein frame
\begin{align}
        \mathcal{L}&=\frac{R}{16\pi G}+\frac{1}{8\pi G}\left(g^{\mu \nu}\partial_\mu \phi \partial_\nu \phi-V(\phi)\right) \nonumber\\
        &-\frac{1}{4}B_{\hat{F}}(\phi)
         \hat{F}^a_{\mu \nu}\hat{F}^{a\mu \nu}
        - B_\psi(\phi) \Bar{\psi}\slashed{D} \psi +\dots\,,
\label{eq:lagrangian}
\end{align}
where $R$ is the Ricci scalar and $\hat{F}$ and $\psi$ are respectively the various gauge field strengths and fermion fields. $D$ are the covariant derivatives including the coupling between fermions and gauge fields. 
In principle the sum extends infinitely over all the massive modes of the string, and they can potentially be coupled. Note that we adopt the notation of previous literature, in which $\phi$ is measured in units of $\sqrt{\hbar \cdot c/(4 \pi G)} = m_\mathrm{pl} / \sqrt{4\pi}$ with the Planck mass $m_\mathrm{pl} \approx 2.176\cdot 10^{-8}\mathrm{kg}$. In particular, the normalization is not the usual $m_\mathrm{pl}/\sqrt{8\pi}$ used in many other contexts in cosmology, leading to slightly unconventional kinetic energy terms in the Lagrangian of \cref{eq:lagrangian} as well as in \cref{eq:field_density,eq:field_pressure} below. We set $\hbar=c=1$.

The field's density and pressure are
\begin{align}
& \rho_\phi = \rho_T +\rho_V = \frac{1}{8\pi G}\left[\dot{\phi}^2 + V(\phi)\right],
\label{eq:field_density}\\   
& P_\phi = P_T+P_V = \frac{1}{8\pi G}\left[\dot{\phi}^2 - V(\phi)\right],
\label{eq:field_pressure}
\end{align}
where $T$ and $V$ denote the kinetic and potential contributions respectively. To these densities, one can associate their corresponding energy density parameters, and their sum $\Omega_\phi = \Omega_T+\Omega_V$.
The dotted quantities are derivatives with respect to the cosmic time $t$, while $\phi^\prime = \frac{d \phi}{d \ln a}$ denotes derivatives with respect to the logarithm of the scale factor, and $\partial_\tau \phi = (aH) \phi^\prime$ derivatives with respect to conformal time $\tau$.

The model's Friedmann and Klein-Gordon equations are 
\begin{subequations}
\begin{align}
    & H^2=\frac{8\pi G}{3} \rho~,\\
    & \ddot{\phi} +3H\dot{\phi}=4\pi G\sigma~,
\label{eq:FieldEquations}
\end{align}
\end{subequations}
where the $\rho$ is the total density of all components of the universe (including the dilaton) and $H=\dot{a}/a$ is the usual Hubble parameter. Furthermore, the interaction of the field is described by
\begin{equation}
    \sigma = \sigma_V + \sigma_m = -\frac{1}{8 \pi G}\frac{\partial V(\phi)}{\partial\phi} + \sum_{i}\alpha_i(\phi)\left(3P_i-\rho_i\right)\,,
\label{eq:sigma}
\end{equation}
whose first term describes the self-interactions of the dilaton from the potential, while the second term describes the dilaton couplings to the other components of the universe.\footnote{Note the perhaps surprising extra factor of $1/2$ in front of the potential derivative in the source term of the Klein-Gordon \cref{eq:FieldEquations}. This is due to the definition we choose for the action of the field's potential in \cref{eq:lagrangian} with an unconventional $(8\pi G)^{-1}$ factor.} The index $i$ spans all components (hadrons, dark matter, radiation ...) with corresponding densities $\rho_i$ and pressures $P_i$\,. The coupling strengths are quantified by coefficients\footnote{Not to be confused with the fine-structure constant $\alpha$ and its value at redshift zero $\alpha(z=0)=\alpha_0$.} $\alpha_i$  given by the logarithmic gradients of their masses
\begin{equation}
\alpha_i(\phi)= \frac{\partial\ln m_i(\phi)}{\partial\phi}~.
\label{eq:couplingsdef}
\end{equation}
This field induced mass variation is a direct signature of the theory of gravity being non-metric. 

As discussed in \cite{DamourPiazzaVeneziano2002}, one can model the $\phi$ dependence of the hadron coupling $\alpha_h$ and dark matter coupling $\alpha_m$ using
\begin{subequations}
\begin{align}
&\alpha_h(\phi) =\alpha_{h,0} e^{-(\phi-\phi_0)}~,\\
&\alpha_m(\phi) = \alpha_{m,0} e^{-(\phi-\phi_0)}~,
\label{eq:alphaharescale}
\end{align}
\end{subequations}
where we introduced the notations $\alpha_{i,0} = \alpha_i (\phi_0)$ and $\phi_0 = \phi(z=0)$. Doing so, the Klein-Gordon equation can be entirely described in terms of the difference $\phi-\phi_0$. The couplings to hadrons/leptons/dark matter are  driving most of the late time cosmological evolution of the field. Another important interaction, albeit more speculative, is that to a model of dark energy (if not generated through the dilaton itself), through a coupling term $\alpha_{\rm DE}$\,. If this component behaves as a cosmological constant, we have $\sigma_{\rm DE} = \alpha_{\rm DE}(3P_{\rm DE} - \rho_{\rm DE}) \sim -4\alpha_{\Lambda}\rho_{\Lambda}$\,. We will only consider the case where $\alpha_{\Lambda}$ is a constant, which was assumed in most of the previous phenomenological studies \cite{Martinelli2015,MartinsVielzeuf2015,MartinsVacher2019} where $\alpha_{\Lambda}$ was denoted $\alpha_{V}$. Note however that this notation was misleading, as this behavior cannot be simply created by some fine tuned potential of $\phi$, but requires some interaction between the dilaton and dark-energy.

The coupling to radiation is always irrelevant, since in that case $\rho_r = 3 P_r$\, and the term of \cref{eq:sigma} always vanishes. The only other interaction of cosmological interest might be that with massive neutrinos, which is left for future work.

It is convenient to treat the contribution from the dilaton potential simply as another species in the $\sigma$ sum, with coupling\footnote{The normalization is set to ensure consistency with the definitions in the literature \cite{DamourPiazzaVeneziano2002}} $\alpha_V = \frac{1}{4}\frac{\partial \ln V }{\partial \phi}$. Note that any constant in the potential $V(\phi) = \Lambda$ adds a term to the Lagrangian \cref{eq:lagrangian} that is effectively equivalent to a cosmological constant. As such, while being conceptually different, the situation in which the runaway dilaton provides the 
source for dark energy with
a constant potential is phenomenologically equivalent to a runaway dilaton field completely decoupled from dark energy ($V=0\,, \alpha_\Lambda=0$) plus a cosmological constant. However, for $\alpha_{\Lambda}$ to be non-zero requires that $V=0$ and $\Lambda$ to be a different source of dark energy. In addition to these two simple scenarios, we will consider the exponential potential
$ V(\phi) = A_xe^{c_x(\phi - \phi_0)}$, leading to $\alpha_V = c_x/4$\, which represents a well motivated potential from string theory \citep{Gasperini2001,DamourPiazzaVeneziano2002}.

\begin{figure}[h]
    \centering
    \includegraphics[width=\columnwidth]{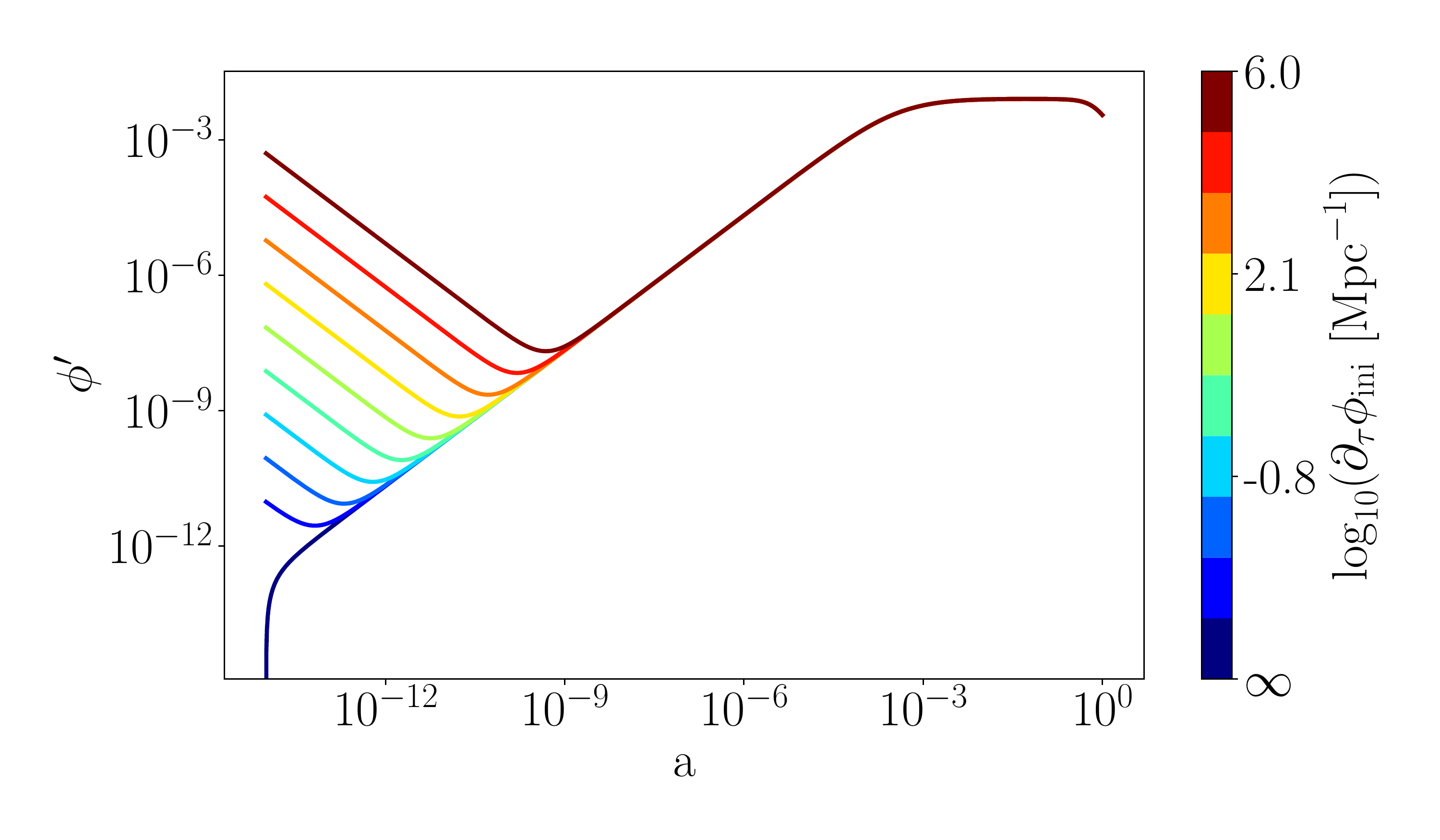}
    \includegraphics[width=\columnwidth]{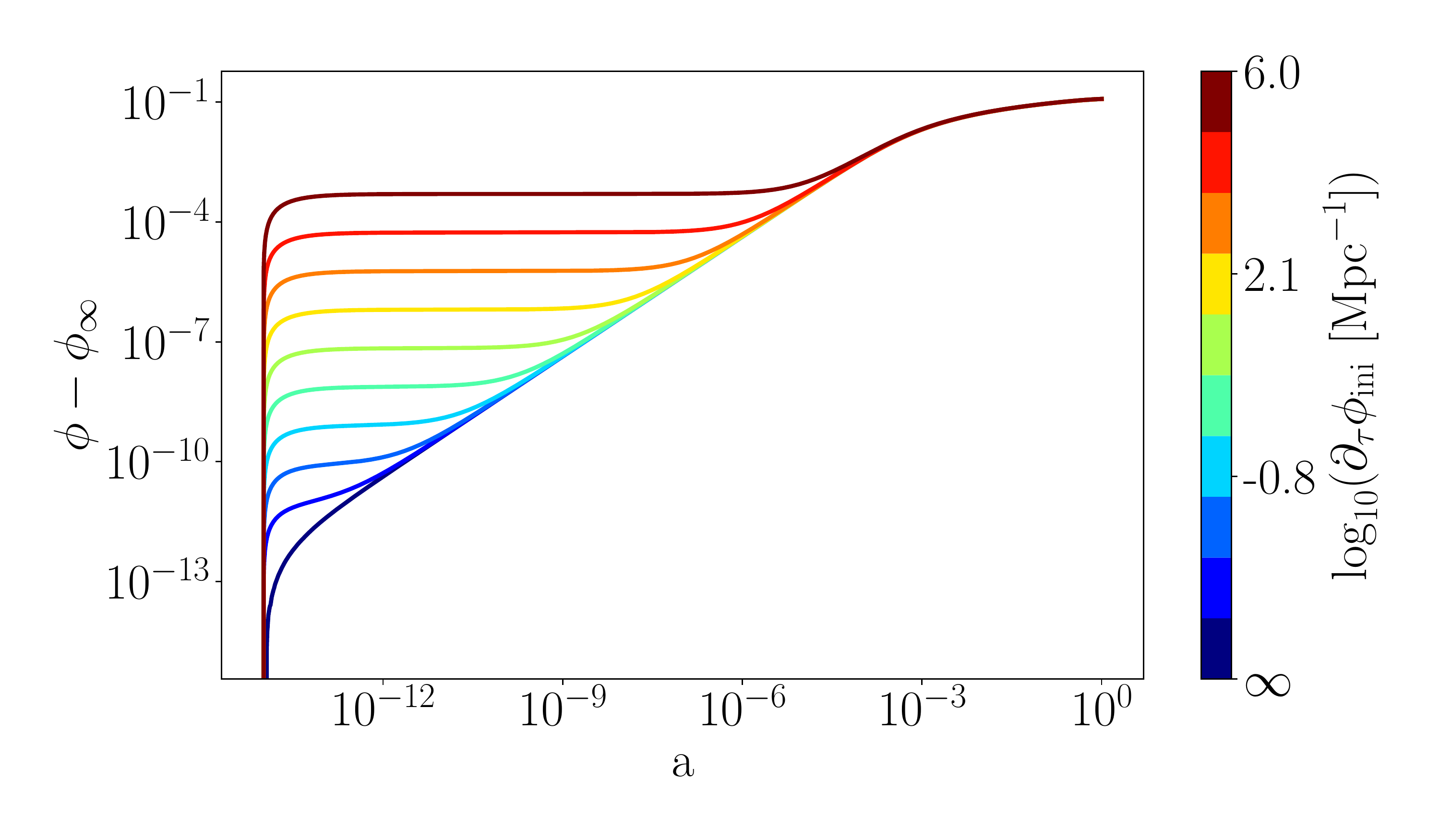}
    \caption{\footnotesize Evolution of the dilaton field and its speed with respect to the scale factor for different values of its initial speed. Here $V(\phi)=0$, $\alpha_{m,0}=-1\times10^{-2}$\,, and $\alpha_{h,0}=-1\times10^{-5}$.}
    \label{fig:attractor}
\end{figure}

The field equations with the couplings as presented thus far display an attractor behavior, shown in \cref{fig:attractor}. First, the initial value of the field is irrelevant in the overall evolution. This is naturally expected from the \cref{eq:FieldEquations,eq:alphaharescale} (which only depend on field differences, not the overall value). Second, there could be, in principle, a dependence on the initial velocity. We observe in \cref{fig:attractor} that due to Hubble friction the field velocity quickly decays from whatever velocity is chosen at the beginning of the evolution to the value that is forced by its interaction with massive species (the ``attractor''). This causes the field $\phi$ to eventually reach a plateau. The overall displacement of the field from its initial value ($\phi-\phi_\infty$) can take on different values at the plateau, depending on the precise initial condition. However, for a large range of initial velocities the late time field velocity (and thus also the overall displacement) is most significant around matter domination, where the acceleration from the coupling is strongest compared to the Hubble friction. In this range the initial velocity is irrelevant.  The starting redshift (here $10^{14}$) is of course set arbitrarily, but this choice does not significantly impact our results.

\subsection{\label{sec:EEPandalpha} Impact on observations}

All the dimensionless coupling coefficients quantifying fundamental interactions of the standard model are expected to be dynamical quantities evolving with the dilaton field itself. The fine-structure constant $\alpha$, quantifying the strength of the electromagnetic interaction, is for this reason expected to exhibit a dynamical behavior and will be directly proportional to the field's coupling to the kinetic term of the Maxwell field strength $F$, $B_F(\phi)$ in the Lagrangian (\cref{eq:lagrangian}). This is particularly relevant due to the extensive astrophysical and laboratory measurements of $\alpha$.

\begin{figure}[t!]
    \centering
    \includegraphics[width=\columnwidth]{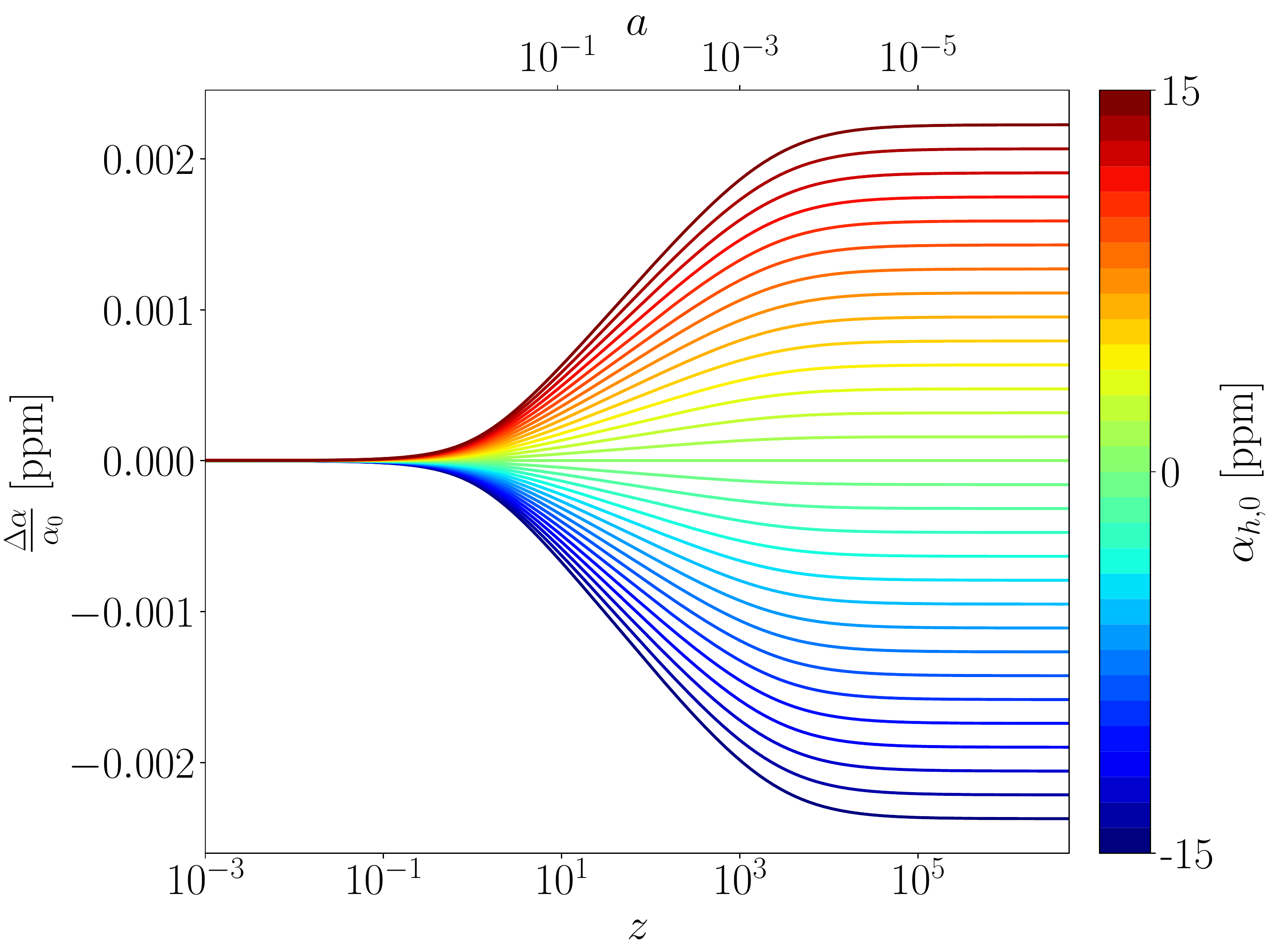}
    \caption{\footnotesize $\frac{\Delta \alpha}{\alpha_0}$ as a function of $z$ and $a$ for different values of $\alpha_{h,0}$. The dark matter coupling is fixed to $\alpha_{m,0}=10^{-3}$ and $\phi_{\rm ini} = \phi'_{\rm ini} = V = 0$.}
    \label{fig:finestrcstz}
\end{figure}

One can  show that the time evolution of $\alpha$ can be linked to the dilaton coupling and field speed as \citep{DamourPiazzaVeneziano2002,MartinsVielzeuf2015}:
\begin{equation}
    \frac{1}{H}\frac{\dot{\alpha}}{\alpha_0} \approx \frac{\alpha_h(\phi)}{40}\phi^\prime~,
\label{eq:dalphadt}
\end{equation}
where $\alpha_0$ is today's value of the fine-structure constant.
This leads to the following redshift dependence
\begin{align}
    &\frac{\Delta \alpha}{\alpha_0}(z) := \frac{\alpha(z)-\alpha_0}{\alpha_0} = \frac{\alpha_{h,0}}{40} \left[ 1 - e^{-(\phi(z)-\phi_0)}\right].
\label{eq:finestrcstz}
\end{align}
An example of this evolution for various dilaton coupling values to hadrons is given in \cref{fig:finestrcstz}.

A different value of $\alpha$ during big bang nucleosynthesis (BBN) also impacts the values of primordial abundances. The most significant of these is the Helium-4 fraction. One can simply model that the induced variation of $Y_{{}^4\mathrm{He}}$ as
\begin{equation}
 \frac{\Delta Y_{^4He}}{Y_{^4He}} = \kappa_{\rm BBN} \frac{\Delta \alpha}{\alpha_0}~.
\label{eq:Helium4BBN}
\end{equation}
For the runaway dilaton, the sensitivity coefficient $\kappa_{\rm BBN}$ is expected to be of order unity \citep{ClaraMartins2020BBN}. We will hence set $\kappa_{\rm BBN}=1$ for the remainder of this work. However, we stress that the impact of this parameter on the analysis is negligibly small.

Furthermore, as discussed in \cite{Planckalphame2016,HartChluba}, $\alpha$ appears in various expressions quantifying the interactions between baryons/leptons with the photons at recombination epoch. 

Ultimately, the atomic energy levels of the Hydrogen atoms are shifted, leading to a delay or advance of recombination. This will impact all the interaction rates and thus, the behavior of the visibility function leading ultimately to a shift of the sound horizon at the last scattering surface, impacting the large $\ell$ values of the angular power spectrum of the cosmic microwave background \cite{Galli:2011,Planckalphame2016} and the value of the Hubble parameter at high redshift \citep{HartChluba,Hart2021,Lee2022}. We self-consistently model this variation of the $\alpha$ fine-structure parameter using \cref{eq:finestrcstz}.

A minor influence on the redshift of reionization is also expected to be induced by a varying $\alpha$. However, the dynamics of reionization is much less known, and the impact would be far less constrained by current data. For this reason, we will ignore it in the present study.

Last but not least, string theory is not a metric theory of gravity, implying that a violation of the Einstein equivalence principle is not only expected but indeed unavoidable at some level \citep{Will2014}.
It can be shown that the Eötvos parameter $\eta$, quantifying deviations from the universality of free fall (UFF) and the Eddington parameter $\gamma$ (related to light deviation by massive objects, and constrained by the Cassini bound) are directly proportional to the square of the dilaton coupling to hadrons \citep{damourEEPscalar,DamourPiazzaVeneziano2002}. At $z=0$, one can derive bounds from general nuclear binding energy formulas
\begin{subequations}
\begin{align}
\label{eq:deviationGR1}
&\eta \simeq 5.2 \times 10^{-5} \alpha_{h,0}^2~,\\
&\gamma-1 \simeq -2\alpha_{h,0}^2~.
\label{eq:deviationGR2}
\end{align}
\end{subequations}

\section{Datasets \label{sec:datasets}}

The runaway dilaton model can be constrained throughout the cosmic evolution using a wide range of local, astrophysical, and cosmological datasets, which we now enumerate.

Local constraints come from experiments on Earth laboratories or in low Earth orbit. Specifically, MICROSCOPE \citep{Microscope_new} provides constraints\footnote{The standard deviation value is obtained by adding quadratically the statistical and systematic errors of \cite{Microscope_new}.} on $\eta$ at $z=0$
\begin{equation}
    \eta = (-1.5 \pm 2.7)\times 10^{-15}\,.
\label{eq:microeta}
\end{equation}

Furthermore, \cite{atomicclock} provides laboratory constraints on the drift rate  $\dot{\alpha}/(\alpha_0 H)$ at $z=0$ using experiments based on atomic clocks, constraining a variation of the fine-structure constant at current times as
\begin{equation}
    \frac{1}{H_0}\left(\frac{\dot{\alpha}}{\alpha_0}\right)_{z=0} = (0.014 \pm 0.015) \times 10^{-6}.
\label{eq:atomicclock}
\end{equation}
Finally, the Oklo natural nuclear reactor \citep{Oklo} provides a geophysical constraint on $\Delta \alpha/\alpha$ 
\begin{equation}
    \frac{\Delta \alpha}{\alpha_0}(z=0.14) = (0.005  \pm 0.061)\times 10^{-6}\,.
\end{equation}
Astrophysical constraints on $\alpha$ are provided by high-resolution spectroscopy of low-density absorption clouds along the line of sight of bright quasars, at low to intermediate redshifts ($z<5$). We used the measurement described in \cite{Martins2017review} combined with recent measurements. All of them can be found in \cite{alphaWebb,alphaSubaru} with an extra point coming from the recent ESPRESSO spectrograph measurement \cite{alphaespresso}. 

Finally, our cosmological data includes {\sc Planck} constraints on CMB power-spectra, lensing \citep{Planck2018,Planck_lkl}, large scale structures and baryon acoustic oscillation from the {\sc BOSS} DR-12 galaxy survey \citep{DR12}. In order to constrain the cosmological background evolution, we will also use the supernovae of type Ia (SNIa) likelihood associated to the {\sc Pantheon} dataset \citep{SNPantheon}. Finally, we also use $H(z)$ measurements coming from recent cosmic-clocks measurements \cite{CC2022}. 

\section{\label{sec:results} Results}

We aim to obtain constraints on the runaway dilaton model free parameters over the whole cosmic history using the datasets presented in \cref{sec:datasets}.

We use a modified version of the {\sc CLASS} software \citep{Lesgourgues:2011CLASS} including the runaway dilaton field. The scalar field impact on background cosmology is computed by integrating the model equations to obtain $\phi(z)$. The code is also modified to consider the various impacts of a redshift dependent value of the fine-structure constant through the cosmic history. In particular, the computed $\Delta \alpha (z) / \alpha_0$ is given by \cref{eq:finestrcstz}. 

In this work, we derive the constraints on the dilaton field simply for the case where the field is spatially homogeneous. However, we have also checked that for cases where the overall energy fraction of the dilaton is subdominant during most of the cosmic evolution, one does not obtain a significant impact of the dilaton field perturbations (when implementing  the usual perturbed Klein Gordon equation, for example). As such, in these cases our results should generalize. Still, we leave a more detailed investigation of the dilaton perturbations for future work.

The likelihood analysis is done by sampling Monte Carlo Markov Chains (MCMC) with {\sc MontePython} \citep{Audren:2012wb,Brinckmann2018} directly coupled to the modified {\sc CLASS} code. We consider the chains to be converged if, for all parameters, the Gelman-Rubin criterion satisfies $|R-1| <
0.05$. Plotting is done using the {\sc Getdist} software \cite{getdist}.

For every run, we sample over the standard cosmological parameters $\{\omega_b, \ln{A_s},n_s,z_{\rm reio},H_0\}$, the dilaton parameters, and the nuisance parameters of the various likelihoods. The priors in all of these parameters are flat and unbounded. In order to remain concise, we will only display the contours for the dilaton parameters most of the time. Note that the values of $\phi_0$ and $\phi'_0$ are derived parameters and not sampled over. While not specified on the figures, their values are always expressed in units of $m_{\rm pl}/\sqrt{4\pi}$\,.

\subsection{Runaway dilaton and a cosmological constant}

In this section we consider the cosmic evolution of a runaway dilaton model decoupled from the cosmological constant, which in this case is the only form of dark energy ($V=0$). This is equivalent to a runaway dilaton with a constant potential $V=\Lambda$ and no cosmological constant. As such, only the dilaton couplings to baryons and/or dark matter are relevant here.

\begin{figure}[h!]
    \centering
    \includegraphics[scale=0.45]{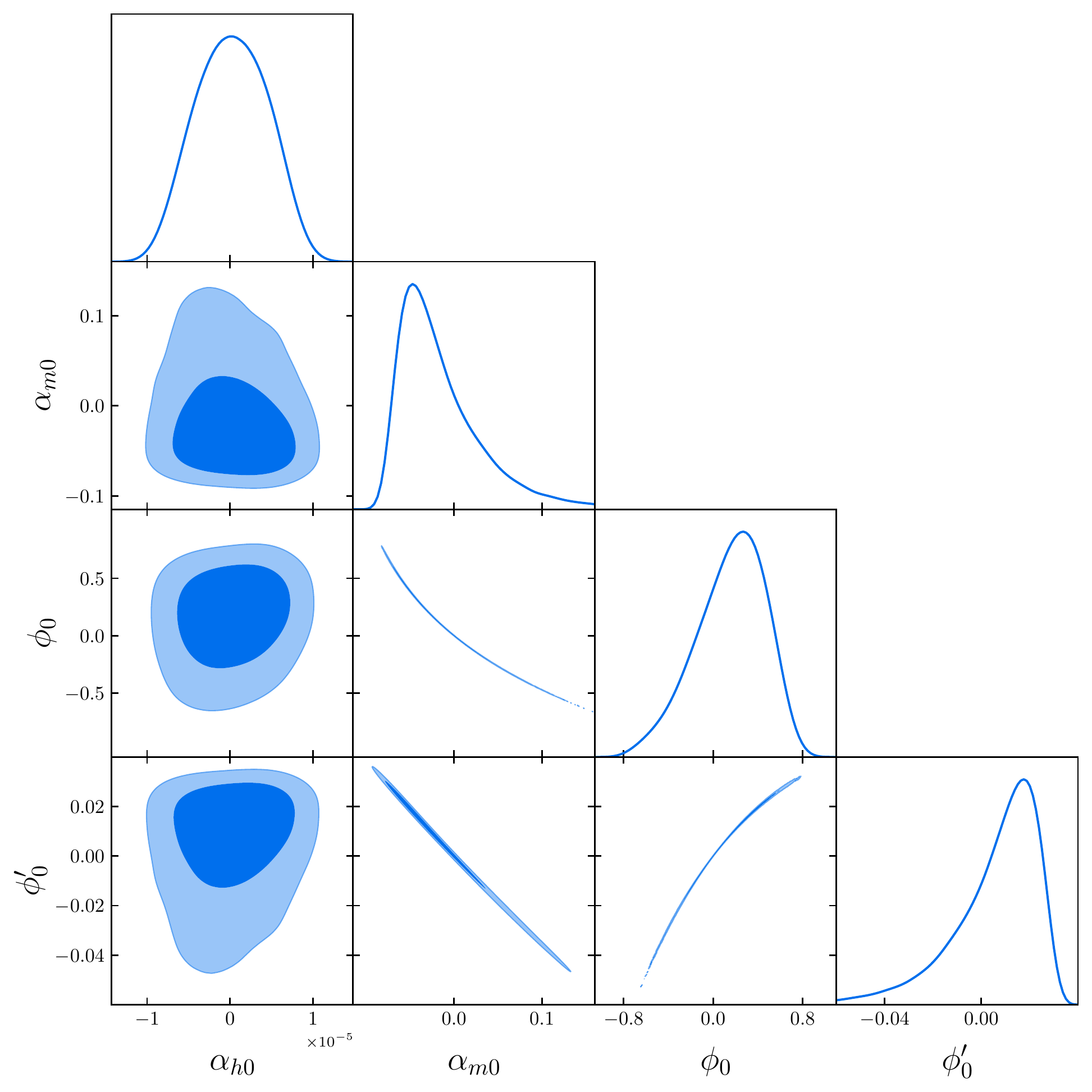}
    \caption{Posteriors of the dilaton parameters with $\alpha_\Lambda =0$ and a constant/zero potential.}
    \label{fig:corneralllkl}
\end{figure}

We display the 68\% and 95\% CL contours of the 2D marginalized posteriors for all combinations of parameters in \cref{fig:corneralllkl} and the corresponding $68\%$ CL are detailed in Tab.~\ref{tab:first}.
 \begin{table}[h!]
    \centering
    \caption{Best-fit values of the runaway dilaton parameters with associated 68$\%$ confidence levels (CL) in the case $V=0$ (or $V=\Lambda$) and $\alpha_{\Lambda}= 0$.}
    \begin{tabular}{l|c}
\toprule            \textbf{Parameter}  & \textbf{68\% CL} \\
\toprule
            {\boldmath$\alpha_{h,0}$} & $(0.24  \,^{+ 4.77}_{-4.57}) \times 10^{-6}$\\
\midrule
                    {\boldmath$\alpha_{m,0}$} & $(-1.33  \,^{+ 1.92 }_{-6.09})\times 10^{-2}$\\

\midrule
                  {\boldmath$\phi_{0}$} & $(1.5^{+4.0}_{-2.4})\times 10^{-1}$\\
\midrule
                  {\boldmath$\phi'_{0}$} & $(5.49  \,^{+ 22.9}_{-7.82}) \times 10^{-3}$ \\
\bottomrule
\end{tabular}
\label{tab:first}
\end{table}
One can witness a very strong correlation between $\alpha_{m,0}$ and today's value of the field $\phi_0$ and its derivative $\phi'_0$\,, while such a correlation is mostly absent with $\alpha_{h,0}$\,. This is expected as the coupling to hadrons is highly constrained by local data as MICROSCOPE while the dark matter coupling, more loosely constrained by the cosmological dataset, has more freedom to accelerate the field toward late times.
Compared to previous studies as \cite{MartinsVacher2019} (which also include $\alpha_\Lambda \neq 0$), the field speed $\phi'_0$ appears however to be more sharply constrained by one order of magnitude, indicating that $\alpha_{m,0}$ does not have an impact on the field evolution as strong as $\alpha_{\Lambda}$ (which here is fixed to 0).

\subsection{Runaway dilaton and a constant coupling to dark energy}

\begin{figure*}[t]
    \includegraphics[scale=0.5]{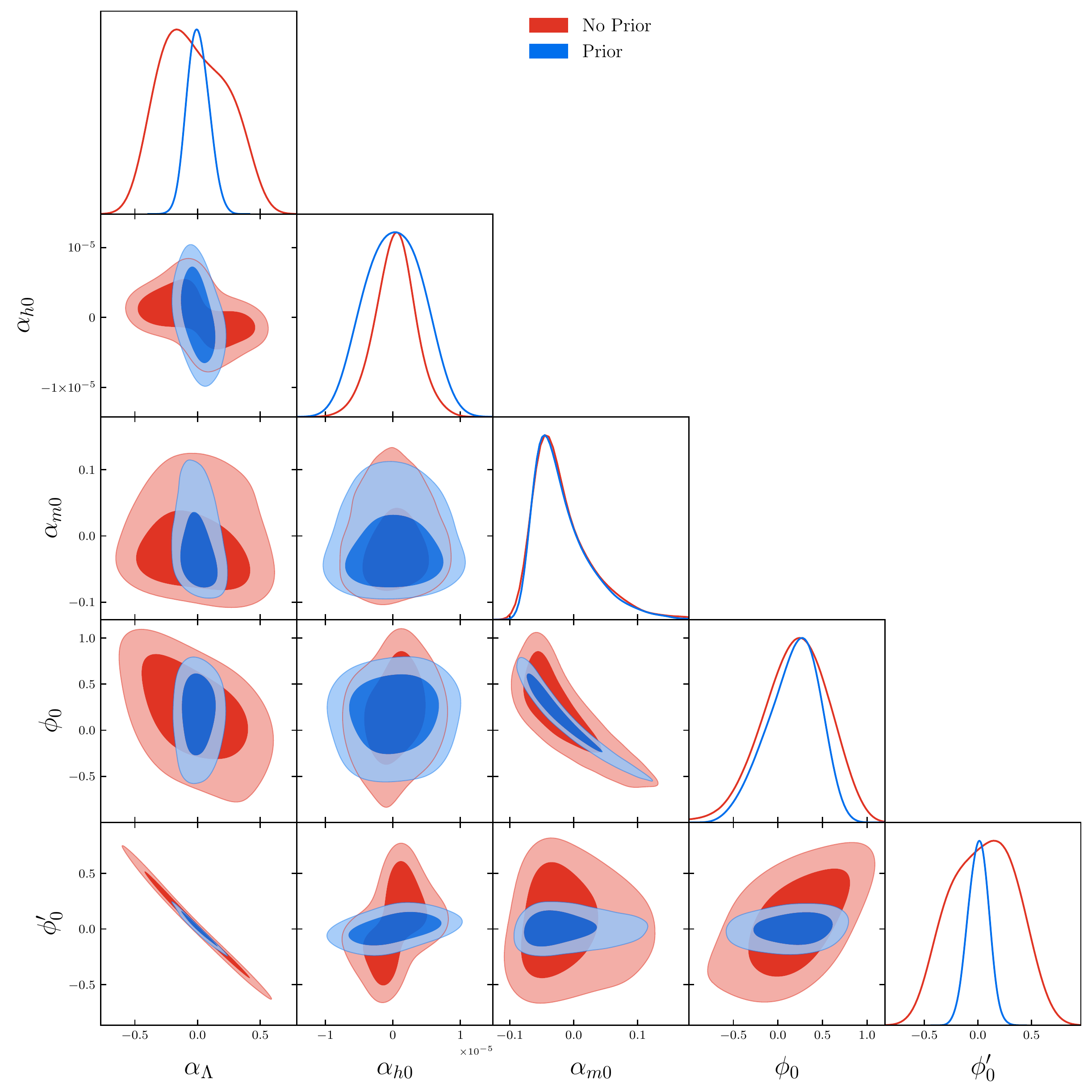}
    \caption{Posteriors of the dilaton parameters with a constant coupling to dark energy $\alpha_{\Lambda}$ with an extra prior on $\phi^\prime_0$ (blue) and without it (red).}
\label{fig:alphalambda}
\end{figure*}

The latest results found in the literature (see e.g. \cite{MartinsVacher2019}) consider the scenario in which $\phi$ can be coupled to $\Lambda$ with a constant coupling. In low redshift studies, an extra prior on today's field speed was given by $|\phi'_0| = 0.0 \pm 0.1$, obtained from separate constraints in \citep{priorphi01,priorphi02}. This prior enables the simplification of the constraints coming from the probes of the cosmological background expansion and therefore provides the main (and effectively the only) constraint on today's field speed $\phi'_0$.
However, using such a prior on today's field speed is in principle unjustified for a full cosmological study since it is derived from rough assumptions (such as matter domination in the current cosmological era), which can be superseded with our likelihood sets. 
\begin{figure*}[t]
\centering
\includegraphics[scale=0.37]{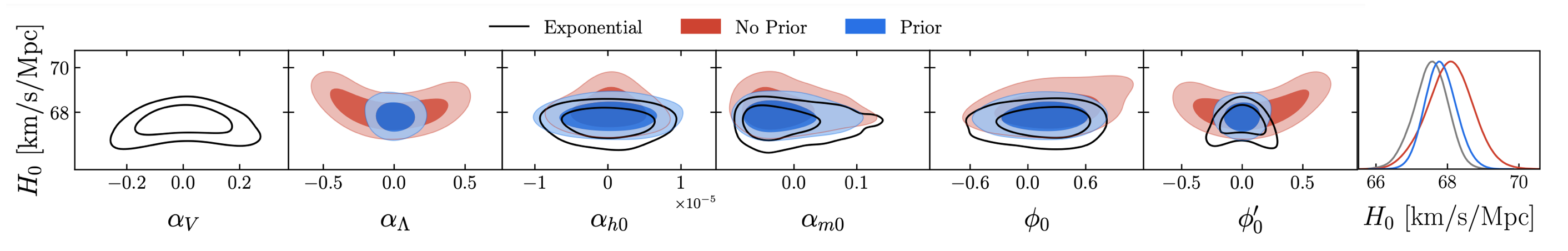} 
\caption{Contour plots of $H_0$ and the dilaton parameters in the cases of an exponential potential (black) and a constant coupling to dark energy $\alpha_{\Lambda}$ with an extra prior on $\phi'_0$ (blue) and without it (red).} \label{fig:H0}
\end{figure*}

We show the results without this prior as the red contours in \cref{fig:alphalambda}, and the results with the prior on $\phi'_0$ as blue contours. We further quantify the results in \cref{tab:second}. These results provide for the first time a study of the full model including $\alpha_{m,0}$ without making any simplifying assumptions (which were called dark, field and matter coupling in the previous studies \citep{MartinsVielzeuf2015,Martinelli2015,MartinsVacher2019}). We find an improvement of the constraints on $\alpha_{h,0}$ by one order of magnitude compared to \cite{MartinsVacher2019}, solely due to the latest MICROSCOPE constraint. The constrains on the coupling to dark energy $\alpha_\Lambda$  are identical when using the prior, as they are an indirect consequence of this restriction set on the field speed, due to the strong degeneracy one can witness between the two parameters. 

While $\alpha_m$ guides the field evolution in matter domination (and thus has a strong impact on the overall field offset $\phi_0$) the impact of the dark energy coupling $\alpha_\Lambda$ is much stronger at late times (around dark energy domination), leading to a very tight degeneracy between $\alpha_\Lambda$ and the current field speed $\phi'_0$\,.

When leaving the prior, the contours are even more non-Gaussian, allowing for large values of $\alpha_{\Lambda}$ and hence of the field speed.
\begin{table}[h]
    \centering
    \caption{Best-fit values of the runaway dilaton parameters with associated 68$\%$ confidence levels (CL) in the case $V=0$ and $\alpha_{\Lambda}\neq 0$.}
    \begin{tabular}{l|c|c}
\toprule            \textbf{Parameter}   & 
            \textbf{Prior on $\phi'_0$} & \textbf{No prior on $\phi'_0$}\\

\midrule
                {\boldmath$\alpha_{h,0}$} & $(-1.63  \,^{+ 4.33 }_{-4.71})\times 10^{-6}$ & $(0.21  \,^{+ 2.97 }_{-2.80})\times 10^{-6}$\\
\midrule
                    {\boldmath$\alpha_{m,0}$} & $(-1.70  \,^{+ 2.08 }_{-5.71})\times 10^{-2}$ & $(-1.39  \,^{+ 2.65}_{-6.03})\times 10^{-2}$\\
\midrule
                  {\boldmath$\alpha_{\Lambda}$} & $(0.50  \,^{+ 8.94 }_{-9.39})\times 10^{-2}$ & $(-0.16\,^{+ 2.34 }_{-3.65})\times 10^{-1}$ \\
\midrule
                  {\boldmath$\phi_{0}$} & $(16.7^{+3.68}_{-2.43})\times 10^{-1}$  & $(17.5^{+4.25}_{-3.23})\times 10^{-1}$ \\
\midrule
                  {\boldmath$\phi'_{0}$}  & $(0.20  \,^{+ 9.97}_{-9.98})\times 10^{-2}$ & $(3.7  \,^{+ 38.4 }_{-31.0})\times 10^{-2}$\\
\bottomrule
\end{tabular}
\label{tab:second}
\end{table}
Surprisingly, the coupling $\alpha_{h,0}$ appears to be $\sim 2$ times more constrained 
without providing any prior on $\phi^\prime_0$, below what the MICROSCOPE bound (\cref{eq:microeta}) can constrain. This is a result from a Bayesian projection effect: The larger space of $\phi^\prime_0$ allowed also allows for a greater amount of models close to $\alpha_{h,0} \sim 0$ to be viable (due to the atomic clock likelihood constraining only the product $\alpha_{h,0}\phi^\prime_0$, see \cref{eq:atomicclock,eq:dalphadt}). This, in turn, explains the specific shape of the contour in the $(\phi'_0,\alpha_{h,0})$ space asking for the two parameters to have the same sign for their product to be positive, and tightens the posterior around $\alpha_{h,0}$ from the Bayesian marginalization. 

\begin{figure}[h!]
    \centering
    \includegraphics[scale=0.6]{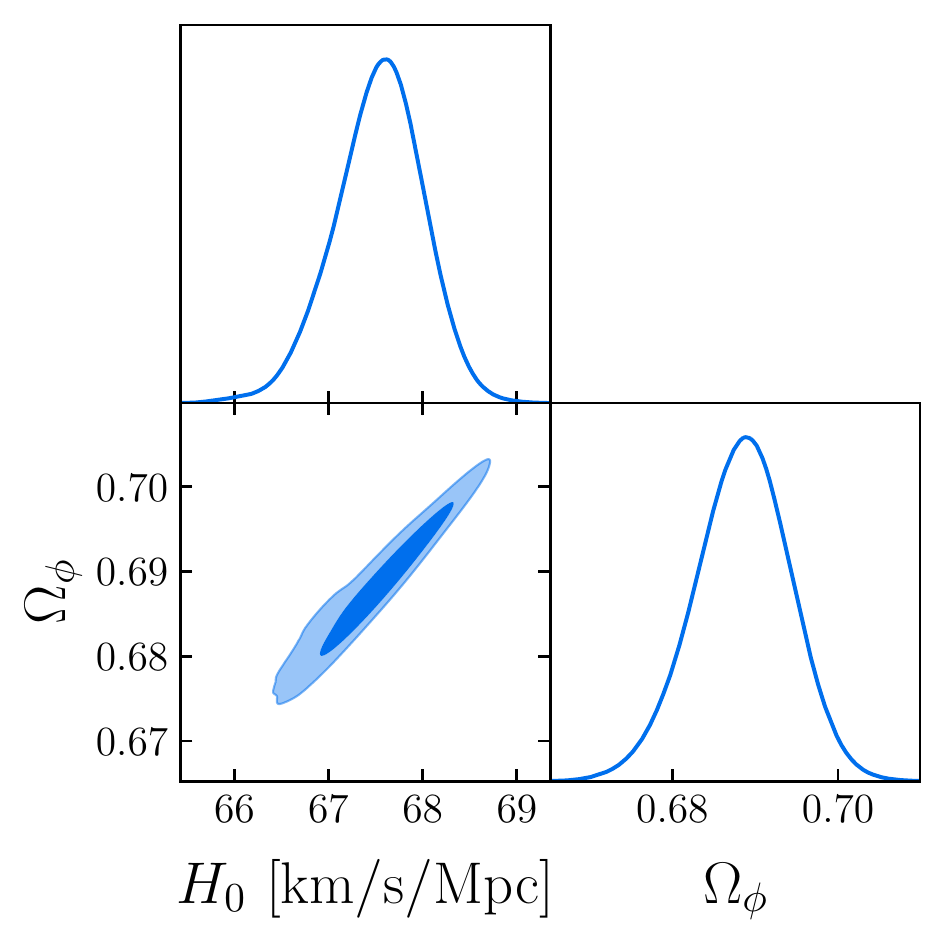}
    \caption{Contour plot of $\Omega_\phi$ and $H_0$ for the exponential potential scenario.}
    \label{fig:Ophi-H0}
\end{figure}

Since we do not have any potential in this case, the overall energy density of the field (\cref{eq:field_density}) is solely given by the kinetic energy of the field ($\Omega_\phi = \Omega_T$). Given that a large coupling to $\Lambda$ is allowed ($|\alpha_\Lambda| \gg 0$), we find that the field strongly accelerates at late times, leading to $\mathrm{d}\rho_\phi/\mathrm{d}\ln a > 0$ (and large $|\phi'_0|$). This naturally allows for a higher $H_0$ due to the geometrical degeneracies in the CMB (compare e.g. with a model of dark energy equation of state with $w<-1$).\footnote{The equation of state of the dilaton naturally always obeys $w_\phi > -1$ since $1+w_\phi = 2 \dot{\phi}^2/[\dot{\phi}^2+V(\phi)] > 0$. However, since we have $\mathrm{d}\rho/\mathrm{d}\ln a > 0$ this is effectively equivalent to a decoupled species with $w< -1$ since for such a species $\mathrm{d}\rho/\mathrm{d} \ln a = -3 (\rho +P) = -3 \rho (1+w) > 0$. The point why such a behavior is preferable can be explained by looking at how late-time solutions to the Hubble tension manage to keep the angular diameter distance (and thus the sound horizon angle) constant. Since we can write $D_A(z_*) \approx \frac{1}{H_0} \int_0^{z_*} \mathrm{d}z/\sqrt{\Omega_\phi(z) + \Omega_\Lambda + \Omega_m (1+z)^3}$, if we increase $H_0$ it is important to decrease the integrand and thus $\Omega_\phi(z)$ in order to keep $D_A(z_*)$ constant. Since $\Omega_\phi(z=0)=1-\Omega_m - \Omega_\Lambda$ is fixed, this can only happen if $\mathrm{d}\Omega_\phi(z)/\mathrm{d}z \propto \mathrm{d}\rho_\phi(z)/\mathrm{d}z\propto -\mathrm{d}\rho_\phi(z)/\mathrm{d}\ln a < 0$.} The contour plots relating the dilaton parameters and $H_0$ are displayed in \cref{fig:H0}. We can observe that today's value of the Hubble parameter $H_0$, is impacted quite strongly by high values of $\phi'_0$\,. Releasing the prior on $\phi'_0$ naturally allows for higher $H_0$ values:
$H_0=68.2_{-0.65}^{+0.51}$\,km/s/Mpc instead of $H_0=67.8 \pm 0.43$\,km/s/Mpc with the prior. 
These conclusions could be relevant in the context of the $\sim 4$-$5 \sigma$ observational tension on the value of $H_0$ and the theoretical limitations of $\Lambda$ as the standard source of dark energy (see e.g.,\citep{cosmoconstreview}). We observe, however, that (due to the atomic clock bound) this quintessence-like behavior of the dilaton field in this configuration is only allowed for smaller values of $\alpha_{h,0}$ and hence smaller violations of general relativity. A targeted and complete study on the role of the dilaton field in this regard remains for future work.

\subsection{Exponential potential}

\begin{figure*}[t!]
    \centering
    \includegraphics[scale=0.5]{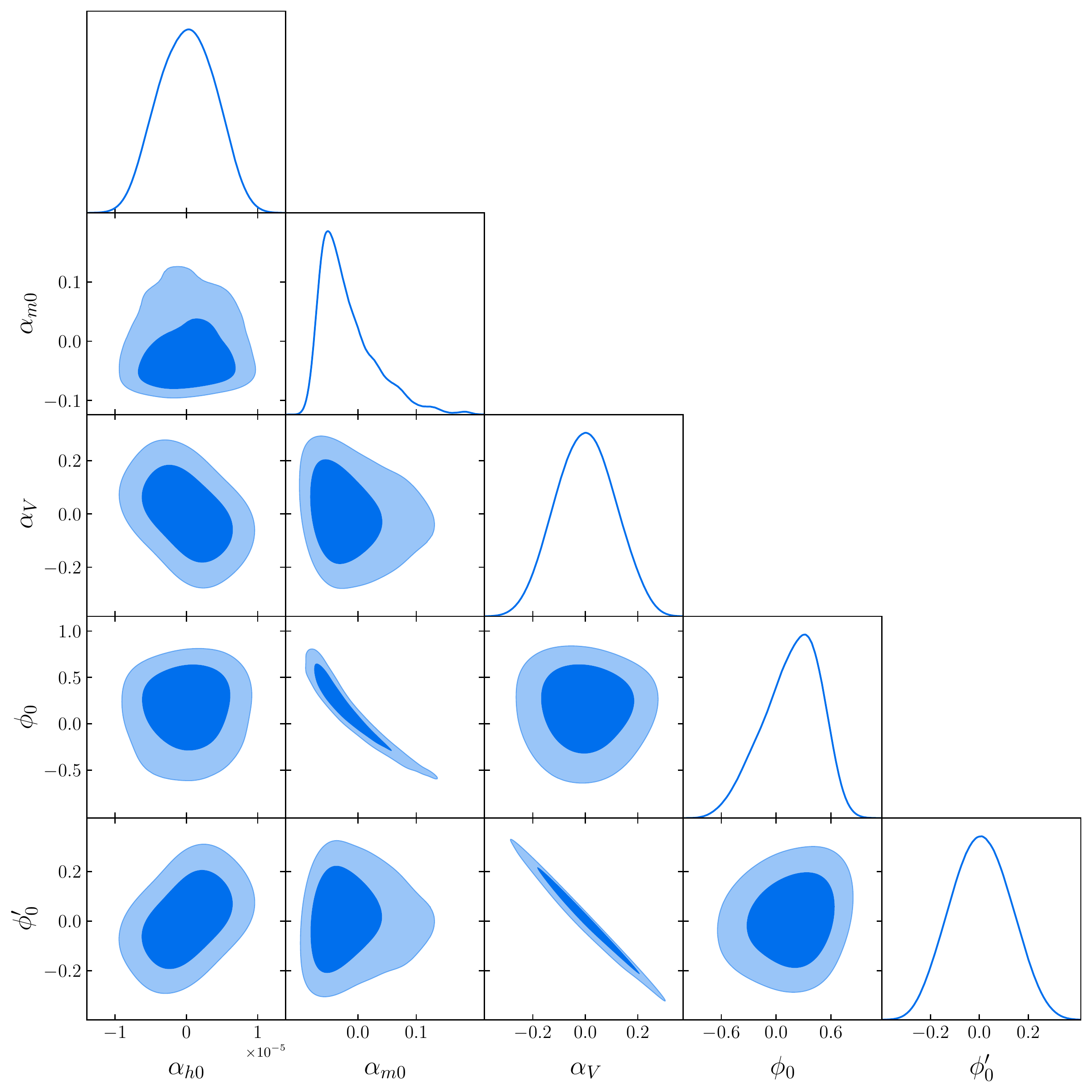}
    \caption{Contour plots for the  dilaton parameters in the exponential potential scenario.}
    \label{fig:exp}
\end{figure*}

We will now consider the case of an exponential shape for $V(\phi)$. In this case the runaway dilaton potential can explain all of the dark energy in the universe, providing that we add a constant term to $V$. The contours are shown in \cref{fig:exp}, and the corresponding constraints are displayed in \cref{tab:third}. As expected, we obtain a high value for $\Omega_\phi = 0.688 \pm 0.006$, showing a strong degeneracy with $H_0$ in \cref{fig:Ophi-H0}. This is expected from the measurement of the CMB sound horizon angle, which tightly constrains $\Omega_m h^3 \approx (1-\Omega_\phi) h^3$. We also observe that this additional degree of freedom does not significantly impact the constraints on $\alpha_{h,0}$ or $\alpha_{m,0}$\,. In this scenario we find that $H_0$ cannot be increased, only decreased. Since the total field energy in this case is dominated by the potential, and one naturally finds $\mathrm{d}V/\mathrm{d}\ln a < 0$ (as long as $|\alpha_V|\gg 0$)\footnote{For a field rolling down its potential one naturally expects $\mathrm{d}V/\mathrm{d}\ln a < 0$, but this can also be confirmed by noticing that $\mathrm{d}V/\mathrm{d}\ln a = \mathrm{d}V/\mathrm{d}\phi \cdot \phi'$ and noticing that due to \cref{eq:FieldEquations} the field speed $\phi'$ naturally evolves in the opposite direction of $\mathrm{d}V/\mathrm{d}\phi$, i.e. $(\phi') ' \propto - \mathrm{d}V/\mathrm{d}\phi$ as long as the Hubble drag and the other coupling terms are comparatively negligible, we also find in this case $\mathrm{d}\rho/\mathrm{d} \ln a < 0$ which (comparably to a dark energy model with $w>-1$) results in lower values of $H_0$\,.}.

{Including both a coupling to a non-negligible cosmological constant and a runaway dilaton potential at the same time causes the parameter space to become extremely hard to sample efficiently. This is because the limit of $\Lambda \to 0$ (with the dilaton potential playing the role of dark energy) naturally allows $\alpha_\Lambda$ to diverge. At the same time, the limit of small $\Omega_\phi$ and correspondingly small $V(\phi)$ also allows the dilaton potential parameters to diverge arbitrarily. As such, instead of imposing arbitrary priors on either the coupling parameters or the cosmological densities, we do not treat this case.}

\begin{table}[h]
    \centering
    \caption{Best-fit values of the runaway dilaton parameters with associated 68$\%$ confidence levels (CL) for the exponential potential case.}
    \begin{tabular}{l|c}
\toprule
            \textbf{Parameter}  & 
            \textbf{68 \% CL}
            \\
\toprule
                {\boldmath$\alpha_{h,0}$} & $(0.01  \,^{+ 4.22 }_{-4.17})\times 10^{-6}$
                \\
\midrule
                    {\boldmath$\alpha_{m,0}$} & $(-1.68 \,^{+ 2.24}_{-5.78})\times 10^{-2}$ 
                    \\
\midrule
                  {\boldmath$\alpha_{V}$} & $(0.04  \,^{+ 1.12}_{-1.27})\times 10^{-1}$ 
                  \\
\midrule
                  {\boldmath$\phi_{0}$} &
                  $(1.64^{+3.82}_{-2.53})\times 10^{-1}$  
                  \\
\midrule
                  {\boldmath$\phi'_{0}$} & $(0.02 \,^{+1.36 }_{-1.26})\times 10^{-1}$
                  \\
\bottomrule
\end{tabular}
\label{tab:third}
\end{table}

\section{Discussion and conclusion} \label{sec:conclusion}

The runaway dilaton model provides a general and self-consistent framework to study the stability of fundamental constants, and the cosmological impact of their space-time variations. It also allows to probe credible models of string theories with existing data-sets. In this work, we obtained the first constraints on the complete parameter space of this model, considering its full cosmological evolution with minimal assumptions on its couplings, updating and refining previous studies. To do so, we benefit from the synergy of multiple independent probes as cosmological, astrophysical, and laboratory datasets. In particular, a major lever arm is provided by the final data release of the MICROSCOPE experiment \cite{Microscope_new}. We explored three scenarios of increasing complexity, showing that order unity couplings (which
would be natural in string theory) are ruled out in all cases. 

While the possible field evolution is expected to be further constrained by the data of incoming wide cosmological surveys as $Euclid$ \cite{Euclid2021}, $DESI$ \cite{Desi2016}, CMB $Stage-4$ \cite{CMBS4} or $LiteBIRD$ \cite{10.1093/ptep/ptac150}, major restriction of its parameter space are expected to be provided by future experiments allowing to directly measure the value of the fine structure constant with an extreme precision, either in laboratory with nuclear clocks \cite{Thorium}, in the nearby universe using spectroscopy \cite{HIRES}, or in the primeval universe with spectral distortions of the CMB \cite{Hart2023}.

Runaway dilaton models (and, more widely, all scalar field induced varying constant models) can additionally play an important role in contemporary debates triggered by the recent discovery of the accelerated expansion of the universe \citep{Riess1998,Perlmutter1999} and the nature of dark energy. As shown in \cite{HartChluba,Hart2021,H0olympics,Lee2022}, a redshift dependence of $\alpha$ -- or possibly of the electron mass $m_e$ -- can have a significant impact on recombination processes that could partially ease or solve the Hubble tension. Providing a suitable choice of couplings or potential, we discussed how the runaway dilaton field can act as dynamical dark energy and significantly impact the value of $H_0$\,. Future studies will reveal if possible extensions of this model can further ease cosmological tensions or if the framework is too restrictive to feasibly do so.

\section*{Acknowledgments}

L.V. would like to thank J. Aumont, A. Blanchard, B. Lamine, J. Lesgourgues and L. Montier who made this collaboration and this work possible by introducing the authors. Extra thanks also goes to S. Nesseris for interesting discussions on {\sc CLASS} and {\sc Montepython} as well as C. M. J. Marques and P. Fayet for feedback on the latest stages of the draft. 
Computations were made on the Mardec cluster supported by the OCEVU Labex (ANR-11-LABX-0060) and the Excellence Initiative of Aix-Marseille University - A*MIDEX, part of the
French “Investissements d’Avenir” program. LV would also like to thanks B. Carreres for several helps with the use of Mardec.

Nils Sch\"oneberg acknowledges the support of the following Maria de Maetzu fellowship grant: Esto publicaci\'on es parte de la ayuda CEX2019-000918-M, financiado por MCIN/AEI/10.13039/501100011033.

This work was financed by Portuguese funds through FCT - Funda\c c\~ao para a Ci\^encia e a Tecnologia in the framework of the project 2022.04048.PTDC. JDFD is supported by an FCT fellowship, grant number SFRH/BD/150990/2021. C.J.M also acknowledges FCT and POCH/FSE (EC) support through Investigador FCT Contract 2021.01214.CEECIND/CP1658/CT0001.

\bibliography{apssamp}
\end{document}